\documentclass[12pt]{article}
\usepackage{amssymb}
\usepackage[dvips]{epsfig}
\usepackage{amsmath,amssymb,indentfirst,epsfig,bm}

\newcommand{\be}{\begin{equation}}
\newcommand{\ee}{\end{equation}}
\def\bea{\begin{eqnarray}}
\def\eea{\end{eqnarray}}

\newcommand{\bn}{\begin{eqnarray}}
\newcommand{\en}{\end{eqnarray}}

\newcommand{\no}{\noindent}

\def\bea{\begin{eqnarray}}
\def\eea{\end{eqnarray}}

\newcommand{\beq}{\begin{eqnarray}}
\newcommand{\eeq}{\end{eqnarray}}

\newcommand{\la}{\mathcal{L}(a_1)}
\newcommand{\lc}{\mathcal{L}_{\mbox{\tiny{nFP}}}(c)}
\newcommand{\bit}{\begin{itemize}}
\newcommand{\eit}{\end{itemize}}

\newcommand{\lag}{\mathcal{L}^g(a_1)}
\newcommand{\lcg}{\mathcal{L}^{g}_{\mbox{\tiny{nFP}}}(c)}
\newcommand{\n}{\nabla}

\newcommand{\ld}{{\mathcal{L}}^{\mbox{\tiny{(dual)}}}}
\newcommand{\lmss}{{\mathcal{L}}^{\mbox{\tiny{(MSS)}}}}

\begin{document}

\title{\textbf{Massive spin-2 particles in a curved background via a nonsymmetric tensor}}
\author{D. Dalmazi\footnote{dalmazi@feg.unesp.br}, H. G. M. Fortes\footnote{hemily.gomes@gmail.com} \\
\textit{{UNESP - Campus de Guaratinguet\'a - DFQ} }\\
\textit{{Avenida Dr. Ariberto Pereira da Cunha, 333} }\\
\textit{{CEP 12516-410 - Guaratinguet\'a - SP - Brazil.} }\\}
\date{\today}
\maketitle

\begin{abstract}
Massive spin-2 particles has been a subject of great interest in
current research. If the graviton has a small mass, the
gravitational force at large distances decreases more rapidly,
which could contribute to explain the accelerated expansion of the
universe. The massive spin-2 particles are commonly described by
the known Fierz-Pauli action which is formulated in terms of a
symmetric tensor $h_{\mu\nu}=h_{\nu\mu}$. However, the Fierz-Pauli
theory is not the only possible description of  massive spin-2
particles via a rank-2 tensor. There are other two families of
models $\mathcal{L}(a_1)$ and $\mathcal{L}_{nFP}(c)$, where $a_1$
and $c$ are real arbitrary parameters, which describe massive
particles of spin-2 in the flat space via a nonsymmetric tensor
$e_{\mu\nu}\neq e_{\nu\mu}$. In the present work we derive
Lagrangian constraints stemming from $\mathcal{L}(a_1)$ and
$\mathcal{L}_{nFP}(c)$ in curved backgrounds with nonminimal
couplings which are analytic functions of $m^2$. We show that the
constraints lead to a correct counting of degrees of freedom if
nonminimal terms are included with fine tuned coefficients and the
background space is of the Einstein type, very much like the
Fierz-Pauli case.  We also examine the existence of local
symmetries.

\end{abstract}

\newpage

\section{Introduction}

Our motivation to work with massive spin-2 particles in a curved
background is twofold. On one hand, they can represent massive
gravitons at the linearized approximation, on the other hand, they
can be understood as elementary massive spin-2 particles in a
given gravitational background.

Regarding the motivation for massive gravitons, they lead to an
weaker gravitational interaction at large distances, which could
contribute to the observed \cite{super1,super2} accelerated
expansion of the universe at large distances. Although the recent
detection of gravitational waves \cite{gw} is consistent with
massless gravitons, predicted by the usual (massless) general
relativity, massive gravitons are not ruled out. The mentioned
experiment sets an upper bound of about $10^{-22} \, eV$ for the
graviton mass \cite{gw2}. Furthermore, previous theoretical
obstacles for massive gravitons like the vDVZ mass discontinuity
\cite{vdv,zak} and the existence of ghosts in the nonlinear theory
\cite{db} have been tackled by the addition of fine tuned
nonlinear self-interaction terms for the metric fluctuation, see
\cite{drgt} and the bimetric model of \cite{hr}. Those models are
based on previous ideas of \cite{vain} and \cite{effective} and
have recently led to intense work on massive gravity and related
topics, see the review works \cite{hinter,drham}.

Regarding elementary massive spin-2 particles, the coupling of
higher spin particles to electromagnetic and gravitational
interactions is a longstanding  problem. Since any elementary
particle must couple to gravity, one first needs to check the
gravitational interaction as in \cite{ad1,ad2} and \cite{bgkp}.
Usually, unitarity \cite{db} and causality \cite{vz,velo} are lost
in interacting theories of higher spin particles. Those particles
require the use of higher rank tensors which have too many
components. The redundant components must vanish on shell. They
work like auxiliary fields. However, when interactions are turned
on, some of those auxiliary fields may become dynamic, giving rise
to negative contributions to the Hamiltonian (instabilities) and
incorrect number of degrees of freedom.

Basically all studies of interacting massive spin-2 particles and
the modern massive gravity theories, as \cite{drgt}, start with
the paradigmatic free theory suggested by Fierz and Pauli in
\cite{FierzPauli}. It describes massive spin-2 particles via a
symmetric and traceful rank-2 tensor $h_{\mu\nu} = h_{\nu\mu}$. It
is the metric fluctuation in massive gravitational theories,
$g_{\mu\nu} = \eta_{\mu\nu} + h_{\mu\nu}$. A natural question
concerns the independence of the outcome of such studies on the
underlying specific massive spin-2 model.

In \cite{Dalmazi2} we have started with a rather general second
order (in derivatives) Ansatz for a quadratic Lagrangian for a
nonsymmetric rank-2 tensor $e_{\mu\nu}$  and by requiring the
existence of only one massive physical pole in the spin-2 sector
of the propagator we have obtained three families of consistent
free theories describing massive spin-2 particles. One of them is
the usual Fierz-Pauli (FP) family which includes the FP model
written in terms of a symmetric tensor. The other two families
require a nonsymmetric tensor. There is no local field
redefinition relating those families in general. One of the
families is given in (\ref{La1}) and the other one in (\ref{Lc}).
They depend on an arbitrary real constant, $a_1$ and $c$,
respectively. See also \cite{Morand} for the special case $a_1 =
-1/4$. Here we couple a background gravitational field to those
theories by including also nonminimal terms and look for curved
space generalization of the tensor, vector and scalar constraints
which are necessary for getting rid of nonphysical degrees of
freedom. We require that the coefficients of the nonmininal terms
be analytic functions of $m^2$. Such restriction plays a key role
in our work and leads us to constraint the gravitational
background to Einstein spaces, see further comments in the
conclusion. In section \ref{sectionLa1} we deal with ${\cal
L}(a_1)$, while in section \ref{sectionlc} we study the ${\cal
L}_{nFP}(c)$ case. In section \ref{concl} we draw our conclusions.
In the appendix we briefly show the technical difficulties in
arbitrary backgrounds.

\section{Family of Lagrangians $\la$}
\label{sectionLa1}

\subsection{Main results in the flat space}

In \cite{Annals} the family of second order Lagrangians $\la$ has
been presented in arbitrary dimensions $D\geq 3$, but here we
focus in $D=4$. It describes massive ``spin-2'' particles via a
nonsymmetric rank-2 tensor $e_{\mu\nu}\neq e_{\nu\mu}$ in the flat
space\footnote{Throughout this work we use
$\eta_{\mu\nu}=(-,+,+,+)$,
$e_{(\alpha\beta)}=(e_{\alpha\beta}+e_{\beta\alpha})/2$ and
$e_{[\alpha\beta]}=(e_{\alpha\beta}-e_{\beta\alpha})/2$.} for any
value of the constant $a_1$:
\begin{eqnarray}
\mathcal{L}(a_1)=-\frac{1}{2}\partial^\mu e^{(\alpha\beta)}\partial_\mu e_{(\alpha\beta)}+\biggl(a_1+\frac{1}{4}\biggr)\partial^\mu e[\partial_\mu e-2\partial^\alpha e_{(\alpha\mu)}]+\nonumber \\
+[\partial^\alpha
e_{(\alpha\beta)}]^2+\biggl(a_1-\frac{1}{4}\biggr)(\partial^\alpha
e_{\alpha\beta})^2-\frac{m^2}{2}(e_{\mu\nu}e^{\nu\mu}- e^2)
\label{La1}
\end{eqnarray}

We recover the FP theory at $a_1=1/4$ where $e_{[\mu\nu]}$ becomes
non dynamic and it can be neglected. However there is no local
field redefinition which takes us from the FP theory to $a_1\neq
1/4$. The massless theory $\mathcal{L}_{m=0}(a_1)$ is unitary in
the ranges $a_1\geq 1/4$ and $a_1\leq -1/12$, it describes
massless spin-2 particles plus a scalar field, except at $a_1=1/4$
and $a_1=-1/12$ where the scalar field disappears. At $a_1=-1/12$
the model $\la$ intersects the nFP (non Fierz-Pauli) family of
section \ref{sectionlc} at $c=-1$, see (\ref{Lc}).

The flat space equations of motion $E_{\mu\nu}\equiv \dfrac{\delta
S(a_1)}{\delta e^{\mu\nu}}=0$ are given by:
\begin{eqnarray}
E_{\mu\nu}&=&\square e_{(\mu\nu)}+2\biggl(a_1+\frac{1}{4}\biggr)[\eta_{\mu\nu}(\partial^\alpha \partial^\beta e_{\alpha\beta}-\square e)+\partial_\mu\partial_\nu e]-\partial_\mu\partial^\alpha e_{(\alpha\nu)}+\nonumber \\
&\, & -\partial_\nu\partial^\alpha
e_{(\alpha\mu)}-2\biggl(a_1-\frac{1}{4}\biggr)\partial_\mu
\partial^\alpha e_{\alpha\nu}+m^2(\eta_{\mu\nu}e-e_{\nu\mu})=0
\end{eqnarray}
From $\partial^\nu E_{\mu\nu}=0$, we have the vector constraint:
\begin{eqnarray}
\partial^\alpha e_{\alpha\nu}=\partial_\nu e
\label{la1eq01}
\end{eqnarray}
Plugging (\ref{la1eq01}) back in $E_{\mu\nu}$ we have, from
$E_{\mu\nu}-E_{\nu\mu}=0$, the tensor constraint:
\begin{eqnarray}
e_{[\mu\nu]}=0 \label{fpc1}
\end{eqnarray}
From (\ref{la1eq01}) and (\ref{fpc1}) back in
$\eta^{\mu\nu}E_{\mu\nu}=0$, we obtain the final scalar
constraint:
\begin{eqnarray}
e=0 \label{fpc2}
\end{eqnarray}
and, consequently, from (\ref{la1eq01}) we have the transverse
condition:
\begin{eqnarray}
\partial^\alpha e_{\alpha\nu}=0
\label{fpc3}
\end{eqnarray}
The equations of motion $E_{\mu\nu}=0$ become the Klein-Gordon
equations
\begin{eqnarray}
(\square -m^2)e_{(\mu\nu)}=0
\end{eqnarray}
The FP conditions (\ref{fpc1}), (\ref{fpc2}) and (\ref{fpc3})
guarantee the correct number of 5 degrees of freedom consistent
with $5=2s+1$, see \cite{giacosa} for a recent derivation of the
FP conditions from first principles.

\subsection{Generalization of $\mathcal{L}(a_1)$ to curved spaces}
\subsubsection{General setup and constraints}

If we want to construct a theory of massive ``spin-2'' field in a
curved space out of a nonsymmetric rank-2 tensor we should provide
the same number of propagating degrees of freedom as in the flat
case. They correspond to the curved space version of the 11
Fierz-Pauli conditions (\ref{fpc1}), (\ref{fpc2}), (\ref{fpc3}),
namely $e_{[\mu\nu]}=0$, $g^{\mu\nu}e_{\mu\nu}=0$ and $\n^\mu
e_{\mu\nu}=0$. Thus, from the 16 components of $e_{\mu\nu}$, we
end up with $16-11=5$ degrees of freedom. Our calculations focus
on the $D=4$ case, but it can be generalized to $D$ dimensions
($D\geq 3$).

Generalizing (\ref{La1}) to curved spacetime we substitute all
derivatives by the covariant ones and add nonminimal terms
containing the curvature tensor as in the FP case \cite{Pershin1}.
They also take care of ordering ambiguities. Requiring a quadratic
theory in derivatives, consistent with the flat limit (\ref{La1})
and at most linear in curvatures, the most general action has the
form\footnote{We disregard nonanalytic functions of $m^2$ and the
term $R_{\alpha\beta\mu\nu}\, e^{\alpha\mu}\, e^{\nu\beta}$ which
is redundant due to the cyclic property
$R_{\mu\nu\alpha\beta}+R_{\mu\alpha\beta\nu}+R_{\mu\beta\nu\alpha}=0$.}
\begin{eqnarray}
\lag &=& -\frac{1}{4}\nabla^\mu e^{\alpha\beta}\,\nabla_\mu e_{\alpha\beta}-\frac{1}{4}\nabla^\mu e^{\alpha\beta}\,\nabla_\mu e_{\beta\alpha}+a_1\nabla^\alpha e_{\alpha\beta}\,\nabla_\mu e^{\mu\beta}+\frac{1}{2}\nabla^\alpha e_{\alpha\beta}\,\nabla_{\mu}e^{\beta\mu}+\nonumber \\
&\,&+\frac{1}{4}\nabla^\alpha
e_{\beta\alpha}\,\nabla_{\mu}e^{\beta\mu}
+\biggl(a_1+\frac{1}{4}\biggr)\nabla^\mu e\,\nabla_\mu e-\biggl(a_1+\frac{1}{4}\biggr)\nabla^\mu e\,\nabla^\alpha \, e_{\alpha\mu}+\nonumber \\
&\,& -\biggl(a_1+\frac{1}{4}\biggr)\nabla^\mu e \,\nabla^{\alpha}e_{\mu\alpha}-\frac{m^2}{2}(e_{\alpha\beta}\,e^{\beta\alpha}-e^2)+f_1\,R\,e^{\alpha\beta}\,e_{\alpha\beta}+f_2\,R\,e^2+\nonumber \\
&\,& +f_3\,R_{\alpha\beta\mu\nu}\,e^{\alpha\mu}\,e^{\beta\nu}+f_4\,R_{\alpha\beta}\,e^{\alpha\mu}\,{e^{\beta}}_\mu +f_5\,R_{\alpha\beta}\,e^{\alpha\beta}\,e+f_6\,R_{\alpha\beta\mu\nu}\,e^{\alpha\beta}\,e^{\mu\nu}+\nonumber \\
&\,& +f_7\,R_{\alpha\beta}\,e^{\alpha\mu}\,{e_\mu}^\beta
+f_8\,R\,e^{\alpha\beta}\,e_{\beta\alpha}+f_9\,R_{\alpha\beta}\,e^{\mu\alpha}\,{e_{\mu}}^\beta
\label{La1g}
\end{eqnarray}
where $f_j$ $(j=1,2,...,9)$ are arbitrary constants for the time
being.

Varying the action with respect to $e^{\rho\sigma}$, we obtain the
equations of motion in curved space: \bea
E_{\rho\sigma}\doteq\frac{\delta S}{\delta e^{\rho\sigma}} &=& \frac{1}{2}\square(e_{\rho\sigma}+e_{\sigma\rho})-2a_1\nabla_{\rho}\nabla^{\mu}e_{\mu\sigma}-\frac{1}{2}\nabla_{\rho}\nabla^\mu e_{\sigma\mu}-\frac{1}{2}\nabla_\sigma\nabla^\mu e_{\mu\rho}-\frac{1}{2}\nabla_\sigma \nabla^\mu e_{\rho\mu}+\nonumber \\
&\,&+2\biggl( a_1+\frac{1}{4}\biggr)\biggl[ -g_{\rho\sigma}\square e +\, \nabla_\rho\nabla_\sigma e +g_{\rho \sigma}\frac{\n^{\mu}\n^\alpha(e_{\alpha\mu}+e_{\mu\alpha})}{2}\biggr]+ \nonumber \\
&\,&-m^2(e_{\sigma\rho}-e\,g_{\rho\sigma})+2f_1\, R\,e_{\rho\sigma}+2f_2\,R\,g_{\rho\sigma}\,e+2f_3\,R_{\rho\beta\sigma\nu}\,e^{\beta\nu}+2f_4\,R_{\rho\beta}\,{e^{\beta}}_\sigma +\nonumber \\
&\,& +f_5\, R_{\rho \sigma}\, e +f_5\,R_{\alpha\beta}\,g_{\rho\sigma}\,e^{\alpha\beta}+2f_6\,R_{\alpha\beta\rho\sigma}\,e^{\alpha\beta}+f_7\,R_{\alpha\sigma}\,{e^{\alpha}}_\rho+f_7\, R_{\rho\alpha}\,{e_\sigma}^\alpha +\nonumber \\
&\,&
+2f_8\,R\,e_{\sigma\rho}+2f_9\,R_{\sigma\beta}\,{e_\rho}^\beta =0
\label{eom} \eea

By applying one derivative on the equations of motion and after
several manipulations, one obtains: \bea
\mathcal{C}_\rho\doteq\n^\sigma E_{\rho\sigma} &=&
+(1-2f_3-2f_6)R_{\rho\lambda\sigma\alpha}\n^\alpha e^{\lambda\sigma}+(1+2f_6)R_{\rho\lambda\sigma\alpha}\n^\alpha e^{\sigma\lambda}+\nonumber \\
&\,& +\biggl(\frac{1}{2}+f_7\biggr) R^{\lambda\alpha}\n_\alpha e_{\lambda\rho}+\biggl(\frac{1}{2}+2f_9\biggr)R^{\lambda\alpha}\n_\alpha e_{\rho\lambda}+(f_7-2a_1)R_{\lambda\rho}\n_\mu e^{\mu\lambda}+\nonumber \\
&\,& +\biggl(2f_4-\frac{1}{2}\biggr)R_{\lambda\rho}\n_\mu e^{\lambda\mu}+2f_1R\n^\sigma e_{\rho\sigma}+\biggl(\frac{1}{4}+2f_1+f_9\biggr)e_{\rho\sigma}\n^\sigma R+\nonumber \\
&\,& +\biggl(2f_2+\frac{f_5}{2}\biggr)e\n_\rho R+(m^2+2f_2R)\n_\rho e+\biggl(\frac{1}{2}+2f_4+2f_6\biggr)e^{\beta\sigma}\n_\sigma R_{\rho\beta}+\nonumber \\
&\,& +\biggl(\frac{1}{2}+2a_1+f_5\biggr)R_{\rho\sigma}\n^\sigma e-(1-2f_3-f_5)e^{\alpha\beta}\n_\rho R_{\alpha\beta}+f_5R_{\alpha\beta}\n_\rho e^{\alpha\beta}+\nonumber \\
&\,& -\biggl(\frac{1}{2}+2f_3+2f_6-f_7\biggr) e^{\sigma\alpha}\n_\sigma R_{\rho\alpha}+\biggl(\frac{1}{4}+\frac{f_7}{2}+2f_8\biggr)e_{\sigma\rho}\n^\sigma R+\nonumber \\
&\,&-(m^2-2f_8R)\n^\sigma e_{\sigma\rho} \label{vv} \eea Now, we
define the tensor $\mathcal{C}_{\rho\sigma}$: \bea
\mathcal{C_{\rho\sigma}} &\doteq & E_{\rho\sigma}-E_{\sigma \rho}=\biggl(-2a_1+\frac{1}{2}\biggr)(\n_\rho\n^\mu e_{\mu\sigma}-\n_\sigma \n^\mu e_{\mu\rho})+\nonumber \\
&\,& +[m^2+2R(f_1-f_8)](e_{\rho\sigma}-e_{\sigma\rho})+2f_3\,R_{\rho\beta\sigma\nu}(e^{\beta\nu}-e^{\nu\beta})+\nonumber \\
&\, &+2f_4(R_{\rho\beta}\, {e^\beta}_\sigma -R_{\sigma\beta}\,
{e^\beta}_\rho )+4f_6\,R_{\alpha\beta\rho\sigma}
\,e^{\alpha\beta}+f_7 {R^\alpha}_\sigma(e_{\alpha\rho}-e_{\rho\alpha})+\nonumber \\
&\,& +f_7
{R_\rho}^\alpha(e_{\sigma\alpha}-e_{\alpha\sigma})+2f_9(R_{\sigma\beta}\,
{e_\rho}^\beta -R_{\rho\beta}\, {e_{\sigma}}^\beta) \label{vt}
\eea In order to find a scalar constraint we have to consider the
most general scalar combination of the equations of motion

 \bea
\mathcal{C}\doteq (b_0\, R\, +b_1m^2)g^{\rho\sigma}E_{\rho\sigma}+b_2
R^{\rho\sigma}E_{\rho\sigma}+b_3\n^\rho\n^\sigma E_{\rho\sigma}
\label{vecompleto} \eea where $b_j$ $(j=0,1,2,3)$ are arbitrary
constants for now. By manipulating and simplifying as much as
possible, we obtain the following expression: \bea
\mathcal{C} = &\,&+[b_3(2f_4+f_7)-b_2]R^{\lambda\alpha}\n_\alpha\n^\rho e_{\lambda \rho}+2b_3(1-f_3)R_{\rho\lambda\sigma\alpha}\,\n^\rho\n^\alpha e^{\lambda\sigma}+\nonumber \\
&\,& -\Biggl[2b_2\biggl(\frac{1}{4}+a_1\biggr)-b_3\biggl(\frac{1}{2}-2a_1+f_7+2f_9\biggr)\Biggr]R^{\lambda\alpha}\n_\alpha \n^\rho e_{\rho\lambda}+\nonumber \\
&\, & +\Biggl[(b_0R+b_1m^2)\biggl(\frac{1}{2}+6a_1\biggr)+2b_2\biggl(\frac{1}{4}+a_1\biggr)R-b_3(m^2-2f_1R-2f_8R)\Biggr]\n^\lambda\n^\rho e_{\rho\lambda}+\nonumber \\
&\,&-\Biggl[(b_0R+b_1m^2)\biggl(\frac{1}{2}+6a_1\biggr) +2b_2\biggl(\frac{1}{4}+a_1\biggr)R -b_3(m^2+2f_2R)\Biggr]\square e+\nonumber \\
&\,&+(b_2+b_3f_5)R_{\lambda\rho}\, \square e^{\lambda \rho}+\Biggl[ b_2+b_3f_5+2\biggl(a_1-\frac{1}{4}\biggl)(b_2+b_3) +b_3\Biggr] R_{\lambda\rho}\n^\lambda\n^\rho e + \mathcal{C}_1\nonumber \\
\label{ve} \eea where $\mathcal{C}_1$ contains up to first
derivatives of $e_{\rho\sigma}$. The expression (\ref{ve}) has
seven terms with second derivatives of $e_{\rho\sigma}$, which
must be eliminated in order to become a scalar constraint. In the
special case of the FP theory ($a_1=1/4$), the last two terms with
second derivatives can only be simultaneously canceled if
$b_2=b_3=0$. Back in the other terms we need $b_0 R + m^2 \,
b_1=0$. However in this case we have no constraint whatsoever.
This is in agreement with \cite{Pershin1} where the authors have
chosen Einstein spaces in order to surmount such difficulty. In
the general case $a_1 \ne 1/4$  we have to find a solution for the
system below:

\begin{eqnarray}
b_3(2f_4+f_7)-b_2=0\\
b_3(1-f_3)=0\\
2b_2\biggl(\frac{1}{4}+a_1\biggr)-b_3\biggl(\frac{1}{2}-2a_1+f_7+2f_9\biggr)=0 \\
(b_0R+b_1m^2)\biggl(\frac{1}{2}+6a_1\biggr)+2b_2\biggl(\frac{1}{4}+a_1\biggr)R-b_3(m^2-2f_1R-2f_8R)=0 \label{systemLa1-1Ap}\\
(b_0R+b_1m^2)\biggl(\frac{1}{2}+6a_1\biggr) +2b_2\biggl(\frac{1}{4}+a_1\biggr)R -b_3(m^2+2f_2R)=0 \label{systemLa1-2Ap}\\
b_2+b_3f_5=0\\
2b_2\biggl(\frac{1}{4}+a_1\biggl)+2b_3\biggl(\frac{1}{4}+a_1+\frac{f_5}{2}\biggl)=0
\label{systemLa1Ap}
\end{eqnarray}

\no Without restrictions in the background, as shown in the
appendix,  we have not been able to solve the previous system and
get $e=0$ from the scalar constraint. So, we are going to restrict
the gravitational background to Einstein
spaces\footnote{Altogether with Bianchi identities we have $\n^\mu
R_{\mu\nu\rho\sigma}=0$ and $\n^\mu R=\partial^\mu R=0$.} as in
the FP case \cite{hinter,Pershin1},

\begin{eqnarray}
R_{\mu\nu}=\frac{R}{4}g_{\mu\nu} \label{es}
\end{eqnarray}

\no Now we can rewrite (\ref{vv}), (\ref{vt})  and (\ref{ve}) as
follows:

\bea \mathcal{C}_\rho\doteq\n^\sigma E_{\rho\sigma} &=&
(1-2f_3-2f_6)R_{\rho\lambda\sigma\alpha}\,\n^\alpha e^{\lambda\sigma}+(1+2f_6)\, R_{\rho\lambda\sigma\alpha}\n^\alpha e^{\sigma\lambda}+2\tilde{f_1}
R\,\n^\lambda e_{\rho\lambda}+\nonumber \\
&\,&+\biggl[\frac{1}{2}\biggl(\frac{1}{4}-a_1+4\tilde{f_8}\biggl)R-m^2\biggr]\n^\lambda e_{\lambda\rho}+\biggl[\frac{1}{2}\biggl(\frac{1}{4}+a_1+4\tilde{f_2}\biggr)R+m^2\biggr]\n_\rho e=0\nonumber \\
\label{vvnovo} \eea \bea
\mathcal{C_{\rho\sigma}}\doteq E_{\rho\sigma}-E_{\sigma \rho}&=&+ \biggl(-2a_1+\frac{1}{2}\biggr)(\n_\rho\n^\mu e_{\mu\sigma}-\n_\sigma \n^\mu e_{\mu\rho})+2(f_3+2f_6)\,R_{\rho\beta\sigma\nu}(e^{\beta\nu}-e^{\nu\beta})+\nonumber \\
&\,&+
\biggl[m^2+(2\tilde{f_1}-2\tilde{f_8})R\biggl](e_{\rho\sigma}-e_{\sigma\rho})=0
\label{vtnovo} \eea \bea
\mathcal{C}& \doteq & \tilde{b}_1 g^{\rho\sigma}E_{\rho\sigma}+b_3\n^\rho\n^\sigma E_{\rho\sigma} \nonumber \\
&=& +2b_3(1-f_3)R_{\rho\lambda\sigma\alpha}\n^\rho\n^\alpha
e^{\lambda\sigma}+
\nonumber \\
&\,&+\Biggl[-6\tilde{b}_1\biggl(a_1+\frac{1}{12}\biggr)+\frac{b_3}{2}\biggl(\frac{1}{4}+a_1+4\tilde{f_2}\biggr)R+b_3m^2\Biggr]\square e+\nonumber \\
&\, & +\Biggl[6\tilde{b}_1\biggl(a_1+\frac{1}{12}\biggr)+\frac{b_3}{2}\biggl(\frac{1}{4}-a_1+4\tilde{f_1}+4\tilde{f_8}\biggr)R-b_3m^2\Biggr]\n^\lambda\n^\rho e_{\rho\lambda}+\nonumber \\
&\,&+\tilde{b}_1\biggl[3m^2+\biggl(2\tilde{f_1}+8\tilde{f_2}+\frac{f_3}{2}+2\tilde{f_8}\biggr)R\biggr]
e =0 \label{venovo} \eea

\no Motivated by the substitution of (\ref{es}) back in
(\ref{La1g}) and (\ref{vecompleto}) we have defined:
\begin{eqnarray}
\tilde{f_1}&\doteq &f_1+\dfrac{f_4}{4}+\dfrac{f_9}{4}\nonumber \\
\tilde{f_2}&\doteq &f_2+\dfrac{f_5}{4}\label{ftilde} \\
\tilde{f_8}&\doteq &f_8+\dfrac{f_7}{4}\nonumber \\ \tilde{b}_1
&\doteq &b_0\, R + b_1 m^2+\dfrac{b_2R}{4}\nonumber
\end{eqnarray}

\no The expression (\ref{vvnovo}) is already a vector constraint
since it does not have second derivatives of the field. It
corresponds to four constraints, in total. The same does not occur
in expressions (\ref{vtnovo}) and (\ref{venovo}). First, let us
turn (\ref{venovo}) into a scalar constraint. We need to solve the
system below: \bea
b_3(1-f_3)=0\nonumber \\
-6\tilde{b}_1\biggl(a_1+\frac{1}{12}\biggr)+\frac{b_3}{2}\biggl(\frac{1}{4}+a_1+4\tilde{f_2}\biggr)R+b_3m^2=0\label{sistema0}\\
6\tilde{b}_1\biggl(a_1+\frac{1}{12}\biggr)+\frac{b_3}{2}\biggl(\frac{1}{4}-a_1+4\tilde{f_1}+4\tilde{f_8}\biggr)R-b_3m^2=0\nonumber
\eea

It is easy to see that the solution of (\ref{sistema0}) back in
(\ref{venovo}) leads to the scalar constraint $e=0$, provided the
coefficient of $e$ is different from zero in (\ref{venovo}).
However, the expression (\ref{vtnovo}) still has terms with second
derivatives. For these terms to be cancelled\footnote{If
$a_1=\dfrac{1}{4}$, those terms would be eliminated, but this
specific value for $a_1$ represents the FP case and it's not of
our interest here.}, it's necessary that $\n^\mu e_{\mu\nu}=0$. We
can get this from the vector constraint (\ref{vvnovo}) if an
appropriate choice of parameters is made. More specifically, since
the solution of (\ref{sistema0}) requires $f_3=1$, if we set
$\tilde{f_1}=0$ and $f_6=-\frac{1}{2}$, we obtain automatically
from (\ref{vvnovo}) that  $\n^\mu e_{\mu\nu}=0$ as far as the
coefficient of $\n^\mu e_{\mu\nu}$ in (\ref{vvnovo}) is non-null.
The solution of the system given in (\ref{sistema0}) with the
additional equations $\tilde{f_1}=0$ and $f_6=-\frac{1}{2}$ is
given by:
\begin{eqnarray}
f_3 &=& 1 \nonumber \\
\tilde{f_8} &=& -\frac{1}{8}-\tilde{f}_2\label{solution1}\\
\tilde{b}_1 &=&
\frac{b_3}{1+12a_1}\Biggl[2m^2+\biggl(\dfrac{1}{4}+a_1+4\tilde{f_2}\biggr)R\Biggr]\nonumber
\end{eqnarray}
Returning this solution in (\ref{vvnovo}), (\ref{vtnovo}) and
(\ref{venovo}) we finally get all necessary constraints. More
specifically, from (\ref{venovo}) we obtain the scalar constraint:
\begin{eqnarray}
e=0 \label{vea1}
\end{eqnarray}
Using (\ref{solution1}) and the result $e=0$ in (\ref{vvnovo}), we
have the vector constraint:
\begin{eqnarray}
\n^\sigma e_{\sigma\rho}=0 \label{vva1}
\end{eqnarray}
Finally, using (\ref{solution1}) and the results (\ref{vea1}) and
(\ref{vva1}) in (\ref{vtnovo}), we achieve the tensor constraint:
\begin{eqnarray}
e_{[\rho\sigma]}=0 \label{vta1}
\end{eqnarray}
once its coefficient in (\ref{vtnovo}) is nonvanishing too.
Summarizing, we have found all the FP constraints:
\begin{eqnarray}
e=0\label{vinculosa1-1}\\
\n^\sigma e_{\sigma\rho}=0\label{vinculosa1-2} \\
e_{[\rho\sigma]}=0 \label{vinculosa1-3}
\end{eqnarray}
if the restriction below are respected
\begin{eqnarray}
b_3\tilde{m}^2\Biggl[2\tilde{m}^2+\biggl(-\dfrac{1}{4}+a_1\biggr)R\Biggr]\Biggl[3\tilde{m}^2-\frac{R}{2}\Biggr]\neq
0 \label{restrictionLa1}
\end{eqnarray}
\begin{eqnarray}
\tilde{m}^2\equiv m^2+\biggl(\frac{1}{4}+2\tilde{f}_2\biggr)R
\label{mtildela1}
\end{eqnarray}
while the equations of motion become
\begin{eqnarray}
E_{\rho\sigma}=(\square
-\tilde{m}^2)e_{\rho\sigma}+2R_{\rho\alpha\sigma\beta}\
e^{\alpha\beta}
\end{eqnarray}

Therefore, we end up with $16-11=5$ degrees of freedom, which is
the correct count for a massive spin-2 particle ($5=2s+1$). The
final curved space theory still contains 2 free parameters:
$\tilde{f_2}$ and $a_1$, with the restriction
(\ref{restrictionLa1}) and $(a_1+1/12)(a_1-1/4)\neq 0$.

For sake of comparison with \cite{Morand} we focus now on a
special subcase of Einstein spaces, namely the maximally symmetric
spaces: \bea
R_{\alpha\beta\rho\sigma}=\frac{R}{12}(g_{\alpha\rho}g_{\beta\sigma}-g_{\alpha\sigma}g_{\beta\rho})
\label{Riems} \eea

All the results from the previous section can be brought
consistently to the maximally symmetric spaces (\ref{Riems}).
Using (\ref{ftilde}) and (\ref{solution1}), the lagrangian
(\ref{La1g}) in a maximally symmetric background becomes \bea
{\mathcal{L}}^{\mbox{\tiny{(MSS)}}}(a_1) &=& -\frac{1}{4}\nabla^\mu e^{\alpha\beta}\,\nabla_\mu e_{\alpha\beta}-\frac{1}{4}\nabla^\mu e^{\alpha\beta}\,\nabla_\mu e_{\beta\alpha}+a_1\nabla^\alpha e_{\alpha\beta}\,\nabla_\mu e^{\mu\beta}+\frac{1}{2}\nabla^\alpha e_{\alpha\beta}\,\nabla_{\mu}e^{\beta\mu}+\nonumber \\
&\,& +\frac{1}{4}\nabla^\alpha
e_{\beta\alpha}\,\nabla_{\mu}e^{\beta\mu}
+\biggl(a_1+\frac{1}{4}\biggr)\nabla^\mu e\,\nabla_\mu e-\biggl(a_1+\frac{1}{4}\biggr)\nabla^\mu e\,(\nabla^\alpha e_{\alpha\mu}+\nabla^\alpha e_{\mu\alpha})+\nonumber \\
&\,&-\frac{m^2}{2}(e_{\alpha\beta}\,e^{\beta\alpha}-e^2)
-\frac{1}{24}\,R\,e^{\alpha\beta}\,e_{\alpha\beta}+\biggl(\tilde{f_2}+\frac{1}{12}\biggr)\,R\,e^2+\nonumber \\
&\,&-\frac{1}{4}\biggl(\frac{11}{12}+{a_1}+4\tilde{f_2}\biggr)\,R\,e^{\alpha\beta}\,e_{\beta\alpha}
\label{lmssa1} \eea On the other hand, it has been presented in
\cite{Morand} a model for massive spin-2 particles also with a
nonsymmetric tensor $e_{\mu\nu}\neq e_{\nu\mu}$ minimally coupled
to maximally symmetric background.

The lagrangian is known as dual massive gravity and is given by
\begin{eqnarray}
\ld =\frac{1}{2}\nabla_\rho e_{\nu\sigma}(-\nabla^\rho
e^{\nu\sigma}-\nabla^\nu e^{\rho\sigma}+\nabla^\nu
e^{\sigma\rho}-\nabla^\rho e^{\sigma\nu}+\nabla^\sigma
e^{\rho\nu}+\nabla^\sigma
e^{\nu\rho})-m^2(e_{\mu\nu}e^{\nu\mu}-e^2)\nonumber \\
\label{acaoW}
\end{eqnarray}

As already discussed in \cite{Dalmazi2}, the model presented in
\cite{Morand} is recovered from $\lmss(a_1)$ in the flat space
when $a_1=-1/4$. However, it's important to notice that the
assumption of $a_1=-1/4$ does not require a maximally symmetric
space as we have shown here.

The relation between $\ld$ and $\lmss(a_1)$ is given by \bea
\lmss(a_1=-1/4)=\frac{1}{2}\ld
-\biggl[\frac{1}{24}+\tilde{f_2}\biggr]R(e^{\alpha\beta}e_{\beta\alpha}-e^2)
\eea Thus, the model of \cite{Morand} is a subcase of
$\lmss(a_1=-1/4)$ where: \bea \tilde{f_2}=-\frac{1}{24} \label{c4}
\eea

\no With the above value of $\tilde{f_2}$, the restrictions
(\ref{restrictionLa1}) lead to two forbidden values for the scalar
curvature, namely, $R \ne -6\, m^2$ and $R \ne 12\, m^2$. The
first value differs by a sign from the restriction obtained in
\cite{Morand} while the second one has not been mentioned. It is
important to emphasize however, that $\tilde{f_2}$ is a free
parameter in the $\lag$ model, so the inequality
(\ref{restrictionLa1}) restricts the possible values of
$\tilde{f_2}$, not of the curvature $R$. This happens because our
original Lagrangian is more general than (\ref{acaoW}). In MSS
there are no forbidden values for the scalar curvature in the
$\lag$ model for any value of $a_1$, including $a_1=-1/4$.

\subsubsection{Local symmetries of $\lag$}

In the previous sections, we have found all the constraints of
$\lag$ model. The form of (\ref{vvnovo}) and (\ref{venovo})
suggests that some local symmetries of $\lag$ may exist even in
the massive case. For example, if the expression obtained for the
vector constraint $\n^\sigma E_{\rho\sigma}$ given in
(\ref{vvnovo}) becomes identically null, instead of a vector
constraint we would have four identities $\n^\sigma
E_{\rho\sigma}\equiv 0$ and, consequently, the theory acquires the
vector symmetry
\begin{eqnarray}
\delta e_{\rho\sigma}=\n_\sigma A_\rho \label{simla1}
\end{eqnarray}
since, up to a surface term,
\begin{eqnarray}
\delta_{\mbox{\tiny{A}}} S=\int d^4x \ \frac{\delta S}{\delta
e^{\rho\sigma}}\ \delta e^{\rho\sigma}=\int d^4 x \ E_{\rho\sigma}
\ \n^\sigma A^\rho =-\int d^4x \ (\n^\sigma E_{\rho\sigma} )A_\rho
=0
\end{eqnarray}
In order that $\n^\sigma E_{\rho\sigma}=0$ holds identically, see
(\ref{vvnovo}), we need the conditions:
\begin{eqnarray}
f_3 &=& 1\nonumber \\
\tilde{f_1}&=&0\nonumber \\
\tilde{f_8}&=&-\frac{1}{8}-\tilde{f_2}\label{condsimla1}\\
f_6&=&-\frac{1}{2}\nonumber \\
R&=&-\frac{8m^2}{1+4a_1+16\tilde{f_2}}\nonumber
\end{eqnarray}

We have checked explicitly that (\ref{simla1}) is indeed a
symmetry of $\lag$ if we use $(\ref{condsimla1})$.

Analogously, we could find scalar symmetries for $\lag$. The
action is invariant under the transformation
\begin{eqnarray}
{\delta e_{\rho\sigma}}^{\mbox{\tiny{(1)}}}=\n_\rho\n_\sigma
\lambda
\end{eqnarray}
where $\lambda$ is an arbitrary scalar, provided the conditions
below are satisfied:
\begin{eqnarray}
f_3&=&1\nonumber \\
\tilde{f_8}&=& -\frac{1}{8}-\tilde{f_1}-\tilde{f_2}\label{cond_sim1} \\
R&=&-\frac{8m^2}{1+4a_1+16\tilde{f_2}}\nonumber
\end{eqnarray}

In addition, there is another possible scalar symmetry which is
\begin{eqnarray}
{\delta e_{\rho\sigma}}^{\mbox{\tiny{(2)}}}=12\, \n_\rho\n_\sigma
\lambda +R\, g_{\rho\sigma}\lambda
\end{eqnarray}
where $\lambda$ is an arbitrary scalar and the conditions below
are required:
\begin{eqnarray}
f_3&=&1\nonumber \\
\tilde{f_8}&=& -\frac{1}{8}-\tilde{f_1}-\tilde{f_2}\label{cond_sim2}\\
R&=&-\frac{12m^2}{1+24\tilde{f_2}}\nonumber
\end{eqnarray}

We leave for a future work a detailed study of the special cases
(\ref{condsimla1}), (\ref{cond_sim1}) and (\ref{cond_sim2}). The
appearance of vector and scalar symmetries are usually connected
with massless and partially massless theories, respectively.

\section{Family of Lagrangians $\lc$ }
\label{sectionlc}
\subsection{Main results in the flat space}

Analogously to the $\la$ case, in \cite{Annals} we can find
another family of second order Lagrangians $\lc$ which describes
massive ``spin-2'' particles via a nonsymmetric rank-2 tensor in
$D=4$ flat spaces\footnote{The model $\lc$ can be generalized to
arbitrary $D\geq 3$, see \cite{Annals}.}
\begin{eqnarray}
\lc=-\frac{1}{2}\partial^\mu e^{(\alpha\beta)}\partial_\mu e_{(\alpha\beta)}+\frac{1}{6}\partial_\mu e[\partial^\mu e-2\partial_\nu e^{(\nu\mu)}]+\nonumber \\
+[\partial^\alpha e_{(\alpha\beta)}]^2-\frac{1}{3}[\partial_\mu
e^{\mu\nu}]^2-\frac{m^2}{2}(e_{\mu\nu}e^{\nu\mu}+c\, e^2)
\label{Lc}
\end{eqnarray}
The real constant $c$ is arbitrary and nFP stands for ``non
Fierz-Pauli'' since we do not need to have $c=-1$. In such special
case, however, the model $\mathcal{L}_{\mbox{\tiny{nFP}}}(c=-1)$
coincides with $\la$ at $a_1=-\frac{1}{12}$. The massless case
$\mathcal{L}^{m=0}_{\mbox{\tiny{nFP}}}(c)$ has first appeared in
\cite{cmu} and describes massless spin-2 particles. If $c\neq
-\frac{1}{4}$ the FP conditions can be derived from the equations
of motion as follows:
\begin{eqnarray}
E_{\mu\nu}&=&\square e_{(\mu\nu)}+\frac{\eta_{\mu\nu}}{3}(\partial^\alpha \partial^\beta e_{\alpha\beta}-\square e)+\frac{1}{3}\partial_\mu\partial_\nu e-\partial_\mu\partial^\alpha e_{(\nu\alpha)}+\nonumber \\
&\, & -\partial_\nu\partial^\alpha
e_{(\alpha\mu)}+\frac{2}{3}\partial_\mu \partial^\alpha
e_{\alpha\nu}-m^2(e_{\nu\mu}+c\, \eta_{\mu\nu}e)=0\label{eomlc}
\end{eqnarray}
From $\partial^\nu E_{\mu\nu}=0$, we have
\begin{eqnarray}
\partial^\nu e_{\nu\mu}+c\, \partial_\mu e=0
\label{lceq01}
\end{eqnarray}
Back in (\ref{eomlc}), we obtain from $E_{\mu\nu}-E_{\nu\mu}=0$:
\begin{eqnarray}
e_{[\mu\nu]}=0 \label{fpc1-1}
\end{eqnarray}
From $\eta^{\mu\nu}E_{\mu\nu}=0$, we have:
\begin{eqnarray}
m^2\biggl(c+\frac{1}{4}\biggr)e=0\Longrightarrow e=0
\label{fpc2-2}
\end{eqnarray}
and, consequently, from (\ref{lceq01}) now we have
\begin{eqnarray}
\partial^\alpha e_{\alpha\nu}=0
\label{fpc3-3}
\end{eqnarray}
Thus, the equations of motion given in (\ref{eomlc}) become the
Klein-Gordon equations:
\begin{eqnarray}
(\square -m^2)e_{(\mu\nu)}=0 \label{eomlc2}
\end{eqnarray}

If $c=-\frac{1}{4}$ the model $\lc$ is invariant under Weyl
transformations:
$\delta_{\mbox{\tiny{W}}}e_{\mu\nu}=\eta_{\mu\nu}\phi$. We can fix
the gauge $e=0$ and obtain all the FP conditions (\ref{fpc1-1}),
(\ref{fpc2-2}) and (\ref{fpc3-3}) and the Klein-Gordon equations
(\ref{eomlc2}).

\subsection{Generalization of ${\mathcal{L}}_{\mbox{\tiny{nFP}}}(c)$ to curved spaces}
\subsubsection{General setup and constraints}

For ${\mathcal{L}}_{\mbox{\tiny{nFP}}}(c)$ the procedure was
analogous to that used for $\la$. The most general expression for
${\mathcal{L}}_{\mbox{\tiny{nFP}}}(c)$ is the following one: \bea
{\mathcal{L}}_{\mbox{\tiny{nFP}}}^g(c)&=&-\frac{1}{4}\n^\mu e^{\alpha\beta}\n_\mu e_{\alpha\beta}-\frac{1}{4}\n^\mu e^{\alpha\beta}\n_\mu e_{\beta\alpha}-\frac{1}{12}\n^\alpha e_{\alpha\beta}\n_\lambda e^{\lambda\beta}+\frac{1}{2}\n^\alpha e_{\alpha\beta}\n_\lambda e^{\beta\lambda}+\nonumber \\
&\,&+\frac{1}{4}\n^\alpha e_{\beta\alpha} \n_\lambda e^{\beta\lambda}+\frac{1}{6}\n^\mu \n_\mu e-\frac{1}{3}\n^\alpha e_{\alpha\beta}\n^\beta e-\frac{m^2}{2}(e_{\alpha\beta}e^{\beta\alpha}+c\, e^2)+\nonumber \\
&\,&+d_1\, R\,e^{\alpha\beta}\,e_{\alpha\beta}+d_2\,R\,e^2+d_3\, R_{\alpha\beta\mu\nu}\,e^{\alpha\mu}\,e^{\beta\nu}+d_4\,R_{\alpha\beta}\,e^{\alpha\mu}\,{e^{\beta}}_\mu+d_5\, R_{\alpha\beta}\,e^{\alpha\beta}\,e+\nonumber \\
&\,&+d_6
\,R_{\alpha\beta\mu\nu}\,e^{\alpha\beta}\,e^{\mu\nu}+d_7\,
R_{\alpha\beta}\,e^{\alpha\mu}\,{e_\mu}^\beta+d_8\,R
\,e^{\alpha\beta}\, e_{\beta\alpha}+d_9
\,R_{\alpha\beta}\,e^{\mu\alpha}\,{e_\mu}^\beta \label{lcg} \eea
where $d_j$ $(j=1,2,...,9)$ are arbitrary constants for now. The
equations of motion are:
\begin{eqnarray}
E_{\rho\sigma}\doteq \dfrac{\delta S_{\mbox{\tiny{nFP}}}^g(c)}{\delta e^{\rho\sigma}}&=&\frac{1}{2}\square (e_{\rho\sigma}+e_{\sigma\rho})+\frac{1}{6}\n_\rho\n^\lambda e_{\lambda\sigma}-\frac{1}{2}(\n_\rho\n^\lambda e_{\sigma\lambda}+\n_\sigma\n^\lambda e_{\rho\lambda})-\frac{1}{2}\n_\sigma \n^\alpha e_{\alpha\rho}+\nonumber \\
&\,& -\frac{1}{3}g_{\rho\sigma}\square e+\frac{1}{3}\n_\rho\n_\sigma e+\frac{1}{3}g_{\rho\sigma}\n^\beta\n^\alpha e_{\alpha\beta}-m^2(e_{\sigma\rho}+c\, e\, g_{\sigma\rho})+2d_1R\,e_{\rho\sigma}+\nonumber\\
&\,&+2d_2\, R\, e\, g_{\rho\sigma}+2d_3\, R_{\rho\beta\sigma\nu}\,e^{\beta\nu}+2d_4\,R_{\rho\beta}\, {e^{\beta}}_\sigma +d_5\, R_{\rho\sigma}\,e +d_5 \, g_{\rho\sigma}\, R^{\alpha\beta}\, e_{\alpha\beta}+\nonumber \\
&\,&+ 2d_6\, R_{\alpha\beta\rho\sigma}\, e^{\alpha\beta}+d_7\, {R^\alpha}_\sigma\, e_{\alpha\rho}+d_7\, {R_\rho}^\alpha e_{\sigma\alpha}+2d_8\, R \, e_{\sigma\rho}+2d_9\, {R_\sigma}^\beta\, e_{\rho\beta}=0\nonumber \\
\end{eqnarray}
The vector constraint is the following expression: \bea
\mathcal{C}_\rho\doteq\n^\sigma E_{\rho\sigma} &=&(1-2d_3-2d_6)R_{\rho\lambda\sigma\alpha}\n^\alpha e^{\lambda\sigma}+(1+2d_6)R_{\rho\lambda\sigma\alpha}\n^\alpha e^{\sigma\lambda}+d_5R_{\alpha\beta}\n_\rho e^{\alpha\beta}+\nonumber \\
&\,&+\biggl(\frac{1}{2}-2d_3-2d_6+d_7\biggr)e^{\lambda\sigma}\n_\lambda R_{\sigma\rho}+(-1+2d_3+d_5)e^{\lambda\sigma}\n_\rho R_{\sigma\lambda}+\nonumber \\
&\,&+\biggl(\frac{1}{2}+2d_4+2d_6\biggr)e^{\sigma\lambda}\n_\lambda R_{\sigma\rho}+\biggl(\frac{1}{2}+2d_9\biggr)R^{\lambda\mu}\n_\mu e_{\rho\lambda}+2d_1R\n^\sigma e_{\rho\sigma}+\nonumber \\
&\,&+\biggl(\frac{1}{2}+d_7\biggr)R^{\lambda\alpha}\n_\alpha
e_{\lambda\rho}+
\biggl(\frac{1}{6}+d_7\biggr)R_{\lambda\rho}\n_\mu e^{\mu\lambda}+\biggl( \frac{1}{4}+2d_1+d_9\biggr)e_{\rho\sigma}\n^\sigma R+\nonumber \\
&\,& +\biggl(2d_4-\frac{1}{2}\biggr)R_{\lambda\rho}\n_\mu
e^{\lambda\mu}+\biggl(\frac{1}{4}+\frac{d_7}{2}+2d_8\biggr)e_{\lambda\rho}\n^\lambda
R+(2d_8R-m^2)\n^\sigma e_{\sigma\rho}+\nonumber
\\
&\,&+\biggl(\frac{1}{3}+d_5\biggr)R_{\alpha\rho}\n^\alpha e
+(2d_2R-m^2c)\n_\rho   e+\biggl(2d_2+\frac{d_5}{2}\biggr) e\n_\rho
R =0 \label{vvlc} \eea The tensor constraint will be obtained from
the expression below: \bea
\mathcal{C}_{\rho\sigma}\doteq E_{\rho\sigma} -E_{\sigma\rho} &=&\frac{2}{3}(\n_\rho\n^\lambda e_{\lambda\sigma}-\n_\sigma\n^\lambda e_{\lambda\rho})+\biggl(m^2+2(d_1-d_8)R\biggr)(e_{\rho\sigma}-e_{\sigma\rho})+\nonumber \\
&\,&+2(2d_6+d_3)R_{\rho\sigma\alpha\beta}\,e^{\alpha\beta}+(2d_4-d_7)({R_\rho}^\beta\, e_{\beta\sigma}-{R_\sigma}^\beta \, e_{\beta\rho})+\nonumber \\
&\,&+(d_7-2d_9)({R_\rho}^\beta\, e_{\sigma\beta}-{R_\sigma}^\beta
\, e_{\rho\beta})=0 \label{vtlc} \eea Regarding the scalar
constraint, due to the Weyl symmetry of the kinetic terms in
$\lcg$ we do not need to add second derivatives of equations of
motion in order to produce a constraint as in (\ref{venovo}). We
can simply have: \bea \mathcal{C} &\doteq &
g^{\rho\sigma}E_{\rho\sigma}\nonumber \\
&=&\biggl[-m^2(1+4c)+(2d_1+8d_2+d_5+2d_8)R\biggr]e+2(d_3+d_4+2d_5+d_7+d_9)R_{\sigma\beta}e^{\sigma\beta}=0\nonumber \\
&\,&
\label{vec-2} \eea

\no Thus, we have a scalar constraint in arbitrary gravitational
backgrounds. In the appendix we show that although, it is possible
to get $e=0$ from (\ref{vec-2}) in arbitrary backgrounds, we are
not able to have a curved space version of the tensor constraint
without restricting the background space. Henceforth we assume
Einstein spaces (\ref{es}). Let us rewrite (\ref{vvlc}),
(\ref{vtlc}) and (\ref{vec-2}) as follows:

\begin{eqnarray}
\mathcal{C}_\rho\doteq\n^{\sigma}E_{\rho\sigma}&=&(1-2d_3-2d_6)R_{\rho\lambda\sigma\mu}\n^{\mu}e^{\lambda\sigma}+(1+2d_6)R_{\rho\lambda\sigma\mu}\n^{\mu}e^{\sigma\lambda} +\tilde{d_1}R\, \n^\sigma e_{\rho\sigma}+\nonumber \\
&\,&+\Biggl[\biggl(\frac{1}{6}+2\tilde{d_8}\biggr)R-m^2\Biggr]\n^{\sigma}e_{\sigma\rho}+\Biggl[\biggl(\frac{1}{12}+2\tilde{d_2}\biggr)R-m^2c\Biggr]\n_{\rho}e=0
\label{vvc}
\end{eqnarray}
\begin{eqnarray}
\mathcal{C}_{\rho\sigma}\doteq  E_{\rho\sigma}-E_{\sigma\rho}&=&\frac{2}{3}(\n_\rho\n^\lambda e_{\lambda\sigma}-\n_\sigma\n^\lambda e_{\lambda\rho})+2(d_3+2d_6)R_{\rho\sigma\alpha\beta}e^{\alpha\beta}+\nonumber \\
&\,&
+\Biggl[m^2+\biggl(2\tilde{d_1}-2\tilde{d_8}+\frac{d_9}{2}\biggr)R\Biggr](e_{\rho\sigma}-e_{\sigma\rho})=0\label{vtc}
\end{eqnarray}
\bea \mathcal{C} &\doteq &
g^{\rho\sigma}E_{\rho\sigma}=\biggl[-m^2(1+4c)+\biggl(2\tilde{d_1}+8\tilde{d_2}+\frac{d_3}{2}+2\tilde{d_8}\biggr)R\biggr]e=0
\label{vec} \eea where we have defined
\begin{eqnarray}
\tilde{d_1}&\doteq &d_1+\dfrac{d_4}{4}+\dfrac{d_9}{4}\nonumber \\ \tilde{d_2}&\doteq &d_2+\dfrac{d_5}{4}\label{dtilde} \\
\tilde{d_8}&\doteq &d_8+\dfrac{d_7}{4}\nonumber
\end{eqnarray}
Therefore, (\ref{vec}) leads to the scalar constraint $e=0$,
provided the coefficient of $e$ is different from zero. On the
other hand, (\ref{vtc}) still has terms with second derivatives.
In order to solve this, we need $\n^\mu e_{\mu\nu}=0$. We can get
this automatically from the vector constraint (\ref{vvc}) by
setting:
\begin{eqnarray}
\tilde{d_1}=0,\hspace{0.5cm} d_3=1, \hspace{0.5cm}
d_6=-\frac{1}{2}
\end{eqnarray}
as far as the coefficient of $\n^\mu e_{\mu\nu}$ does not vanish.

Back in the tensor constraint (\ref{vtc}) we obtain
$e_{[\mu\nu]}=0$ as far as its coefficient is nonvanishing too.
In summary, all 11 Fierz-Pauli constraints (\ref{vinculosa1-1}),
(\ref{vinculosa1-2}) and (\ref{vinculosa1-3}) are confirmed if
\begin{eqnarray}
\tilde{m}^2\biggl(\tilde{m}^2-\frac{R}{6}\biggr)\Biggl\{(1+4c)\tilde{m}^2+\biggl[8(c\,
\tilde{d}_8-\tilde{d}_2)-\frac{1}{2}\biggr]R\Biggr\}\neq 0
\label{condmtil}
\end{eqnarray}
\begin{eqnarray}
\tilde{m}^2\equiv m^2-2\tilde{d}_8R
\end{eqnarray}
while the equations of motion become:
\begin{eqnarray}
E_{\rho\sigma}=(\square -\tilde{m}^2)\,
e_{\rho\sigma}+2R_{\rho\alpha\sigma\beta}\ e^{\alpha\beta}
\end{eqnarray}
where the free paramenters $\tilde{d_2}$ and $\tilde{d_8}$ must
satisfy the conditions (\ref{condmtil}).


\subsubsection{Local symmetries of $\lcg$}

The model $\lcg$ also presents vector and scalar symmetries. There
is one vector symmetry which comes from the transformation
\begin{eqnarray}
\delta e_{\rho\sigma}=\n_\sigma A_\rho
\end{eqnarray}
where $A_\rho$ is an arbitrary vector. We need a nonvanishing
scalar curvature ($R\neq 0$) and the following conditions:
\begin{eqnarray}
\tilde{d_1}&=&0; \hspace{0.5cm} d_6=-\frac{1}{2}; \hspace{0.5cm}d_3=1;\nonumber \\
\tilde{m}^2&=&\frac{R}{6}\\
\tilde{d_2}&=&-\frac{1}{24}+\frac{c}{12}+\tilde{d_8}c\nonumber
\end{eqnarray}
Such conditions imply that the vector constraint (\ref{vvc}) be
identically null, i.e., $\n^\sigma E_{\rho\sigma}\equiv 0$.

On the other hand, starting from the general scalar transformation
\begin{eqnarray}
\delta e_{\rho\sigma}=A_1 \n_\rho \n_\sigma \lambda +A_2 \,
g_{\rho\sigma}\, \lambda +A_3\,  g_{\rho\sigma}\, \square \,
\lambda
\end{eqnarray}
where $\lambda$ is an arbitrary field and $A_j$ $(j=1,2,3)$ are
arbitrary constants, we have found three scalar symmetries for
$\lcg$.

The first one  is
\begin{eqnarray}
{\delta e_{\rho\sigma}}^{\mbox{\tiny{(1)}}}=g_{\rho\sigma}\
\lambda \label{symm1}
\end{eqnarray}
where we must have
\begin{eqnarray}
(1+4c)\tilde{m}^2=R\biggl[\frac{1}{2}+8(\tilde{d}_2-c\,
\tilde{d}_8)\biggr] \label{cond_sim3}
\end{eqnarray}
The second scalar symmetry is
\begin{eqnarray}
{\delta e_{\rho\sigma}}^{\mbox{\tiny{(2)}}}=-4\n_\rho\n_\sigma
\lambda+g_{\rho\sigma}\, \square \lambda \label{symm2}
\end{eqnarray}
where the relations below must hold:
\begin{eqnarray}
d_3&=&1\nonumber \\
\tilde{m}^2&=&(1+12\tilde{d}_1)\frac{R}{6} \label{cond_sim4}
\end{eqnarray}
Finally, the last scalar symmetry is given by:
\begin{eqnarray}
{\delta
e_{\rho\sigma}}^{\mbox{\tiny{(3)}}}=\n_\rho\n_\sigma\lambda
\label{symm3}
\end{eqnarray}
where the conditions below must be obeyed:
\begin{eqnarray}
d_3&=&1\nonumber \\
\tilde{d_2}&=&-\frac{1}{24}+\frac{c}{12}+c(\tilde{d_1}+\tilde{d_8})\label{condsymm3} \\
\tilde{m}^2&=&(1+12\tilde{d}_1)\frac{R}{6}\nonumber
\label{cond_sim5}
\end{eqnarray}
The symmetry (\ref{symm3}) and conditions (\ref{condsymm3}) follow
from the previous scalar symmetries (\ref{symm1}) and
(\ref{symm2}).

\section{Conclusion}
\label{concl}

Here we have studied massive spin-2 models via a nonsymmetric
rank-2 tensor in a curved background. As in the usual Fierz-Pauli
(FP) case with a symmetric tensor, nonminimal couplings are
necessary. The work here is a preliminary one and parallels the
work \cite{Pershin1} on the FP model. As in that case we have
assumed that the Ans\"{a}tze (\ref{La1g}) and (\ref{lcg}) are
linear on curvatures and their coefficients are analytic functions
of $m^2$. Although there seems to be slightly more freedom now in
choosing the background metric than in the FP case, we have
selected, for simplicity,  background spaces of the Einstein type
as in \cite{Pershin1}. We have succeeded in finding nontrivial
solutions for the coefficients of our
 Ans\"{a}tze by getting rid of second derivatives in the tensor,
 vector and scalar constraints. In particular, we have generalized
 from maximally symmetric spaces to Einstein spaces a previous work in the literature \cite{Morand}
 carried out for a massive spin-2 theory with nonsymmetric tensor $e_{\mu\nu} \ne e_{\nu\mu}$, which corresponds
 to the model ${\cal L}(a_1)$ of section 2 at the specific point
 $a_1=-1/4$. Regarding the model ${\cal L}_{nFP}(c)$, due to the
 Weyl symmetry of the kinetic terms, the scalar constraint has
 easily led to the traceless condition $e=0$ but now the problem
 moved to the tensor constraint and once again we have found
 convenient to choose Einstein spaces.

 Comparing the results obtained here in Einstein spaces for ${\cal L}(a_1)$ and
 ${\cal L}_{nFP}(c)$ with the ones obtained in \cite{Pershin1} for the usual FP
 model, the main difference is that, besides the scalar and vector
 constraint, we now have a tensor constraint $C_{\rho\mu}=0$, see
 (\ref{vt}) and (\ref{vtnovo}). However, due to the vector
 constraint $\nabla^{\mu}e_{\mu\nu} =0$, the constraint
 $C_{\rho\mu}=0$ amounts to $e_{\mu\nu}- e_{\nu\mu} =0$ without further restriction in the background. So there
 seems to be no fundamental difference to the usual FP case in
 curved space. Since in the FP case there is no restriction on the
 background metric when  we allow the coefficients to be non-analytic
 functions of $m^2$, we would like to address that point also
 in the case of our nonsymmetric models. This is under
 investigation now.

 Moreover, in both cases of ${\cal L}(a_1)$ and ${\cal
 L}_{nFP}(c)$, we believe that less restrictive conditions on the coefficients can be obtained by
 getting rid of second order derivatives of time only, this is under
 study. We are also analyzing the special points in the parameters space
 where the local symmetries mentioned in the previous sections show up. They indicate
massless and partially massless theories even if $m^2
 \ne 0$. The truly massless cases $m=0$ in both ${\cal L}(a_1)$ and ${\cal
 L}_{nFP}(c)$ theories are also worth investigating in curved space. Especially
 in the second case where, at least in the flat space, we have
massless spin-2 particles just like in the massless version of the
FP model.

As a final comment, we notice that the ghost free massive gravity
theories, see \cite{drgt} and \cite{hr}, accommodate massive
gravitons propagating in any gravitational background, see
\cite{new1,BDvS,new2}. Those results agree with earlier
perturbative (in powers of $1/m^2$ ) calculations. Thus, if we
obtain correct Lagrangian constraints for the models discussed
here, in the case of nonanalytic coefficients, we would be
prompted to search for nonlinear (self-interacting) versions of
${\cal L}(a_1)$ and ${\cal L}_{nFP}(c)$.

\section{Acknowledgements}

H. G. M. F. thanks FAPESP (2013/25368-0)  and D. D. thanks CNPq
(307278/2013-1) for financial support.

\section{Appendix - Constraints in a general background}

\subsection{${\cal L}(a_1)$}

We are looking for a solution of the system (14)-(20) without
imposing restrictions on the background space. Since $f_i$ $
(i=1,..,9)$ and $b_j$ $(j=0,..,3)$ are constants, we demand that
the coefficients of $R$ in the equations (\ref{systemLa1-1Ap}) and
(\ref{systemLa1-2Ap}) are null, i.e.,
\begin{eqnarray}
b_0\biggl(\frac{1}{2}+6a_1\biggr)+2b_2\biggl(\frac{1}{4}+a_1\biggr)+2b_3(f_1+f_8)=0\\
b_0\biggl(\frac{1}{2}+6a_1\biggr)+2b_2\biggl(\frac{1}{4}+a_1\biggr)-2b_3f_2=0
\quad .
\end{eqnarray}
The solution, singular at $a_1 = 1/4$ as expected, is given by
\begin{eqnarray}
f_3&=&1\nonumber \\
f_4&=&\frac{1}{2}+f_9\nonumber \\
f_5&=&\frac{(1+4a_1)}{(-1+4a_1)}\nonumber \\
f_7&=&-\frac{8a_1}{(-1+4a_1)}-2f_9\\
f_2&=&-\frac{1+8a_1(1+2a_1)}{4(-1+4a_1)}+\frac{b_0}{2b_1}\nonumber \\
f_1&=& \frac{1+8a_1(1+2a_1)}{4(-1+4a_1)}-\frac{b_0}{2b_1}-f_8\nonumber \\
b_3&=&\frac{1}{2}\, b_1(1+12a_1)\nonumber \\
b_2&=&-\frac{b_1(1+12a_1)(1+4a_1)}{2(-1+4a_1)}\nonumber
\label{system}\end{eqnarray} Plugging this solution back in
(\ref{ve}), we obtain a scalar contraint in a general background:
\begin{eqnarray}
\mathcal{C}&=&\mathcal{C}_1=\frac{(1+12a_1)}{(-1+4a_1)}\Biggl\{-\frac{b_1(1+4a_1)^2}{2(-1+4a_1)}R^{\rho\sigma}R_{\rho\sigma} e+\frac{b_1(1-12a_1)}{2}\n^\lambda R^{\sigma\rho}\n_\rho e_{\lambda\sigma}+\nonumber \\
&\,&+4b_1a_1e_{\lambda\sigma}\square R^{\lambda\sigma}+\frac{b_1(1+12a_1)}{2(-1+4a_1)}\n_\rho R^{\lambda\sigma}\n^\rho e_{\lambda\sigma}-\frac{b_1(1+4a_1)}{2}\n^\sigma R^{\lambda\alpha}\n_\alpha e_{\lambda\sigma}\Biggr\}+\nonumber \\
&\,&+\frac{b_1(1+12a_1)}{2}\Biggl\{\biggl[-\frac{b_0}{b_1}+\frac{1+8a_1(1+2a_1)}{4(-1+4a_1)}\biggr]\n^\lambda R\, \n^\rho e_{\rho\lambda}+\nonumber \\
&\,&-\biggl[\frac{b_0}{b_1}+\frac{1-2a_1(-1+4a_1)}{-1+4a_1}-2f_9\biggr]\n^\lambda R\, \n^\rho e_{\lambda\rho}\Biggr\}+\nonumber \\
&\,&+\frac{(1+12a_1)}{(-1+4a_1)}\Biggl\{(b_0R+b_1m^2)+\frac{b_1(1+4a_1)}{2}\biggl[m^2+\frac{b_0R}{b_1}-\frac{[16a_1(1+a_1)+3]R}{2(-1+4a_1)}\biggr]\Biggr\}R^{\rho\sigma}e_{\rho\sigma}+\nonumber \\
&\,&+b_1(1+12a_1)R^{\rho\sigma}R_{\rho\beta\sigma\nu}e^{\beta\nu}+\frac{(1+12a_1)}{(-1+4a_1)}\biggl[\frac{b_1(1+12a_1)(1+4a_1)}{2(-1+4a_1)}-8b_1a_1\biggr]R^{\rho\sigma}R_{\rho\beta}\, {e^{\beta}}_\sigma +\nonumber \\
&\,&+\frac{b_1(1+12a_1)}{2}\Biggl\{\biggl[\frac{2b_0}{b_1}-\frac{1+16a_1(1+3a_1)}{4(-1+4a_1)}\biggr]\n_\lambda R \, \n^\lambda e+\biggl[\frac{b_0}{b_1}-\frac{2a_1(1+4a_1)}{(-1+4a_1)}\biggr] (\square R)\, e\Biggr\}+\nonumber \\
&\,&+\Biggl\{(b_0R+b_1m^2)\biggl[3m^2-\frac{1+16a_1(1+3a_1)}{2(-1+4a_1)}R+\frac{3b_0R}{b_1}\biggr]+\nonumber \\
&\,&-\frac{b_1(1+12a_1)(1+4a_1)}{2(-1+4a_1)}\biggr[m^2+\frac{b_0R}{b_1}-\frac{[1+8a_1(1+2a_1)]R}{2(-1+4a_1)}\biggr]R\Biggr\}e\label{82}
\end{eqnarray}

\no We have not been able to derive $e=0$ from (\ref{82}) without
restrictions on the background space.

\subsection{${\cal L}_{nFP}$}

For arbitrary backgrounds if we choose $d_9=-d_3-d_4-2d_5-d_7$ and
$d_2 = - (2d_1+d_5+2d_8)/8$  and assume $c \ne -1/4$, the scalar
constraint (\ref{vec-2}) becomes simply $e=0$. Putting those
results back in (\ref{vvlc}) and (\ref{vtlc}), we obtain:

\begin{eqnarray}
\mathcal{C}_\rho &=& (1-2d_3-2d_6)R_{\rho\lambda\sigma\alpha}\n^\alpha e^{\lambda\sigma}+(1+2d_6)R_{\rho\lambda\sigma\alpha}\n^\alpha e^{\sigma\lambda}+d_5R_{\alpha\beta}\n_\rho e^{\alpha\beta}+\nonumber \\
&\,&+\biggl(\frac{1}{2}-2d_3-2d_6+d_7\biggr)e^{\lambda\sigma}\n_\lambda R_{\sigma\rho}+(-1+2d_3+d_5)e^{\lambda\sigma}\n_\rho R_{\sigma\lambda}+\nonumber \\
&\,&+\biggl(\frac{1}{2}+2d_4+2d_6\biggr)e^{\sigma\lambda}\n_\lambda R_{\sigma\rho}+\biggl(\frac{1}{2}-2d_3-2d_4-4d_5-2d_7\biggr)R^{\lambda\mu}\n_\mu e_{\rho\lambda}+\nonumber \\
&\,&+2d_1R\n^\sigma e_{\rho\sigma}+\biggl( \frac{1}{4}+2d_1-d_3-d_4-2d_5-d_7\biggr)e_{\rho\sigma}\n^\sigma R+\nonumber \\
&\,& +\biggl(2d_4-\frac{1}{2}\biggr)R_{\lambda\rho}\n_\mu
e^{\lambda\mu}+\biggl(\frac{1}{4}+\frac{d_7}{2}+2d_8\biggr)e_{\lambda\rho}\n^\lambda
R+(2d_8R-m^2)\n^\sigma e_{\sigma\rho}+\nonumber \\
&\,&+\biggl(\frac{1}{2}+d_7\biggr)R^{\lambda\alpha}\n_\alpha
e_{\lambda\rho}+
\biggl(\frac{1}{6}+d_7\biggr)R_{\lambda\rho}\n_\mu e^{\mu\lambda}
\label{vvlc-2}
\end{eqnarray}
\bea
\mathcal{C}_{\rho\sigma}&=&\frac{2}{3}(\n_\rho\n^\lambda e_{\lambda\sigma}-\n_\sigma\n^\lambda e_{\lambda\rho})+\biggl(m^2+2(d_1-d_8)R\biggr)(e_{\rho\sigma}-e_{\sigma\rho})+\nonumber \\
&\,&+2(2d_6+d_3)R_{\rho\sigma\alpha\beta}\,e^{\alpha\beta}+(2d_4-d_7)({R_\rho}^\beta\, e_{\beta\sigma}-{R_\sigma}^\beta \, e_{\beta\rho})+\nonumber \\
&\,&+(2d_3+2d_4+4d_5+3d_7)({R_\rho}^\beta\,
e_{\sigma\beta}-{R_\sigma}^\beta \, e_{\rho\beta})=0
\label{vtlc-2} \eea

\no There are still second derivatives in
$\mathcal{C}_{\rho\sigma}$. We could think of determining
$\nabla^{\sigma}e_{\sigma\rho}$ as a function of the remaining
terms in (\ref{vvlc-2}) at $d_8=0$ and plugging it back in
(\ref{vtlc-2}) which leads to:

\bea
\mathcal{C_{\rho\sigma}}&=&\frac{4}{3m^2}\Biggl[(1-2d_3-2d_6)R_{[\sigma\lambda\beta\mu}\n_{\rho]}\n^\mu e^{\lambda\beta}+(1+2d_6)R_{[\sigma\lambda\beta\mu}\n_{\rho]}\n^\mu e^{\beta\lambda}+\nonumber \\
&\,&+\biggl(\frac{1}{2}-2d_3-2d_4-4d_5-2d_7\biggr)R^{\lambda\mu}\n_{[\rho}\n_\mu e_{\sigma]\lambda}+\biggl(\frac{1}{2}+d_7\biggr)R^{\lambda\mu}\n_{[\rho}\n_\mu e_{\lambda\sigma]}+\nonumber \\
&\,& +\biggl(\frac{1}{6}+d_7\biggr)R_{\mu[\sigma}\n_{\rho]}\n_\lambda e^{\lambda\mu}+\biggl(-\frac{1}{2}+2d_4\biggr)R_{\mu[\sigma}\n_{\rho]}\n_\lambda e^{\mu\lambda}+d_5R_{\alpha\beta}\n_{[\rho}\n_{\sigma]} e^{\alpha\beta}+\nonumber \\
&\,&+2d_1R\n_{[\rho}\n^\beta e_{\sigma]\beta}+\n_{[\rho} \mathcal{F}_{\sigma]}\Biggr]+\biggl(m^2+2(d_1-d_8)R\biggr)(e_{\rho\sigma}-e_{\sigma\rho})+\nonumber \\
&\,& +2(2d_6+d_3)R_{\rho\sigma\alpha\beta}\,e^{\alpha\beta}+(2d_4-d_7)({R_\rho}^\beta\, e_{\beta\sigma}-{R_\sigma}^\beta \, e_{\beta\rho})+\nonumber \\
&\,&+(2d_3+2d_4+4d_5+3d_7)({R_\rho}^\beta\,
e_{\sigma\beta}-{R_\sigma}^\beta \, e_{\rho\beta})
 \label{lastf} \eea
where $\mathcal{F}_\alpha$ does not contain any derivative of
$e_{\rho\sigma}$ and is defined as follows: \bea
\mathcal{F}_\alpha &\doteq &+\biggl(\frac{1}{2}-2d_3-2d_6+d_7\biggr)e^{\lambda\mu}\n_\lambda R_{\mu\alpha}+(-1+2d_3+d_5)e^{\lambda\mu}\n_\alpha R_{\mu\lambda}+\nonumber \\
&\,&+\biggl(\frac{1}{2}+2d_4+2d_6\biggr)e^{\mu\lambda}\n_\lambda R_{\mu\alpha}+\biggl(\frac{1}{4}+2d_1-d_3-d_4-2d_5-d_7\biggr)e_{\alpha\lambda}\n^\lambda R+\nonumber \\
&\,&
+\biggl(\frac{1}{4}+\frac{d_7}{2}\biggr)e_{\lambda\alpha}\n^\lambda
R \eea

\no Unfortunately, without any restriction on the background space
we have not been able to avoid second derivatives of $e_{\mu\nu}$
in the tensor constraint (\ref{lastf}). Therefore, just like the
$\lag$ case, we are led to Einstein spaces
$R_{\mu\nu}=\frac{R}{4}g_{\mu\nu}$ once again.

\end{document}